\documentclass{aa}
\usepackage[varg]{txfonts}

\usepackage{hyperref}
\usepackage{array, multirow, graphicx,natbib,txfonts,color,tabularx,amsmath}
\usepackage{lscape,float,placeins}
\usepackage{hyperref}
\usepackage{afterpage}
\usepackage{threeparttable}
\bibpunct{(}{)}{;}{a}{}{,}

\newcommand{\FeH}{$\left[\mbox{Fe/H}\right]$}

\newcommand{\kms}{km\,s$^{-1}$}
\newcommand{\Teff}{$T_{\mbox{{\scriptsize eff}}}$}

\newcommand{\abund}[1]{\log\,\varepsilon(\text{#1})}
\newcommand{\ch}[1]{\multicolumn 1 c {#1}}
\begin{document}

\title{Atomic diffusion and mixing in old stars \\ VI: The lithium content of M30 \thanks{Based on data collected at the ESO telescopes under program 085.D-0375.}\,\thanks{Tables 1 and 5 are only available in electronic form
at the CDS via anonymous ftp to \url{cdsarc.u-strasbg.fr} (130.79.128.5)
or via \url{http://cdsweb.u-strasbg.fr/cgi-bin/qcat?J/A+A/} and the figures are available in colour in the electronic form.} }

\author{Pieter Gruyters\inst{1,2} \and Karin Lind\inst{1}  \and Olivier Richard\inst{3} \and Frank Grundahl\inst{4} \and Martin Asplund\inst{5} \and Luca Casagrande\inst{5} \and Corinne Charbonnel\inst{6,7} \and Antonino Milone\inst{5} \and Francesca Primas\inst{8} \and Andreas J. Korn\inst{1} }

\offprints{pieter.gruyters@physics.uu.se}

\institute{Division of Astronomy and Space Physics, Department of Physics and Astronomy, Uppsala University, Box 516, 75120 Uppsala, Sweden  \and Lund Observatory, Box 43, 221 00 Lund, Sweden \and LUPM, Universit\'e de Montpellier, CNRS, CC072, Place E. Bataillon, 34095 Montpellier Cedex, France \and Stellar Astrophysics Centre, Department of Physics and Astronomy, {\AA}rhus University, Ny Munkegade 120, DK-8000 {\AA}rhus C, Denmark \and Research School of Astronomy and Astrophysics, Mount Stromlo Observatory, The Australian National University, ACT 2611, Australia \and Department of Astronomy, University of Geneva, Chemin des Maillettes 51, 1290 Versoix, Switzerland \and IRAP, CNRS UMR 5277, Universit\'e de Toulouse, 14, Av. E. Belin, 31400 Toulouse, France \and European Southern Observatory, Garching, Germany}

\date{Received  / Accepted}

\authorrunning{P. Gruyters}
\titlerunning{Atomic Diffusion in M30}

\abstract
{The prediction of the PLANCK-constrained primordial lithium abundance in the Universe is in discordance with the observed Li abundances in warm Population II dwarf and subgiant stars. Among the physically best motivated ideas, it has been suggested that this discrepancy can be alleviated if the stars observed today had undergone photospheric depletion of lithium.}
{The cause of this depletion is investigated by accurately tracing the behaviour of the lithium abundances as a function of effective temperature. Globular clusters are ideal laboratories for such an abundance analysis as the relative stellar parameters of their stars can be precisely determined.
}
{We performed a homogeneous chemical abundance analysis of 144 stars in the metal-poor globular cluster M30, ranging from the cluster turnoff point to the tip of the red giant branch. Non-local
thermal equilibrium (NLTE) abundances for Li, Ca, and Fe were derived where possible by fitting spectra obtained with VLT/FLAMES-GIRAFFE using the quantitative-spectroscopy package SME. Stellar parameters were derived by matching isochrones to the observed $V$ vs $V-I$ colour-magnitude diagram. Independent effective temperatures were obtained from automated profile fitting of the Balmer lines and by applying colour-\Teff\ calibrations to the broadband photometry. 
}
{Li abundances of the turnoff and early subgiant stars form a thin plateau that is broken off abruptly in the middle of the SGB as a result of the onset of Li dilution caused by the first dredge-up. 
Abundance trends with effective temperature for Fe and Ca are observed and compared to predictions from stellar structure models including atomic diffusion and ad hoc additional mixing below the surface convection zone. The comparison shows that the stars in M30 are affected by atomic diffusion and additional mixing, but we were unable to determine the efficiency of the additional mixing precisely. This is the fourth globular cluster (after NGC\,6397, NGC\,6752, and M4) in which atomic diffusion signatures are detected. After applying a conservative correction (T6.0 model) for atomic diffusion, we find an initial Li abundance of $A(\text{Li}) = 2.48\pm0.10$ for the globular cluster M30.\\
We also detected a Li-rich SGB star with a Li abundance of $A(\text{Li}) = 2.39$. The finding makes Li-rich mass transfer a likely scenario for this star and rules out models in which its Li enhancement is created during the RGB bump phase.
}
{}

\keywords{stars: abundances - stars: atmospheres - stars: fundamental parameters - globular clusters: individual: M30 - techniques: spectroscopic }

\maketitle


\section{Introduction}\label{sect:intro}
The primordial abundance of lithium is traced from observations of warm metal-poor turnoff stars in the halo of our Galaxy. The lithium abundance in these Population II stars (Pop II) with a metallicity range $-3.0<$ \FeH \footnote{We adopt here the customary spectroscopic notations that [X/Y] $\equiv$ log\,(N$_X$/N$_Y$)$_{*}$--log\,(N$_X$/N$_Y$)$_{\odot}$, and that log\,$\varepsilon(X)\equiv$ log\,(N$_X$/N$_{\rm H}$)+12 for elements $X$ and $Y$.}$<-1.0$  is found to be almost independent of metallicity, displaying a plateau \citep{Spite1982}.
Although lithium in surface layers of stars can be destroyed as a result of convective motions -- surface material can be dragged into the hot stellar interior, where lithium is readily destroyed by proton capture -- lithium is not or scarcely affected in the hottest (\Teff$>6000$\,K) and most massive unevolved halo stars since these have only a thin convection layer. For these stars, lithium shows no correlation with temperature. This is the Spite plateau.
If one assumes that lithium is not being destroyed during the pre-main sequence phase and it is not depleted at the surface of the stars so that the currently observed abundance is equal to the initial one, the Spite plateau level ought to measure the primordial abundance of lithium. The small scatter around the Spite plateau can be interpreted as an indication that depletion of lithium cannot have been very effective and did not vary much from star to star. Several groups \citep[e.g.][]{Ryan1999,Charbonnel2005,Asplund2006,Bonifacio2007, Aoki2009, Hosford2010, Melendez2010} have undertaken great effort to measure lithium abundances of halo stars to create sufficient samples where the dominant errors are systematic in origin. These efforts have led to a value for the Spite plateau of $2.20\pm0.09$ \citep{Sbordone2010}. This is significantly lower than the prediction for the initial Li abundance of the Universe of $N(^7$Li)/$N$(H)$ = (4.68\pm0.67)\times10^{-10}$ or $A$(Li)$=2.66\pm0.06$\footnote{$A$(Li)$ = \log{\frac{N\text{(Li)}}{N\text{(H)}}}+12$.}, obtained from standard Big Bang nucleosynthesis (BBN) based on the most recent determination of the baryon density from the PLANCK data $\Omega_{\text{b}}h^2 = 0.022305\pm0.000225$ \citep{Cyburt2015}. \\ 

During the past decade, advances in theoretical modelling of stellar structure were made. Where the canonical models do not predict Li destruction during the main-sequence lifetime of a Pop II star, stellar structure models by \citet{Richard2005} based on the work by \citet{Michaud1984} revealed that the low Li values found in old, metal-poor stars can be naturally explained. 

\citet{Richard2005} showed that the low Li abundances observed in stars are a result of Li depletion by atomic diffusion in competition with an additional transport or mixing process (hereafter AddMix) in the radiative zones of Pop II stars. By using sophisticated stellar models that treat atomic diffusion including radiative acceleration from first principles and calibrated on the Sun, and assuming rather strict limits on the extent and efficiency of AddMix, \citet{Richard2005} found they could produce a Li plateau for Pop II stars that is $\sim0.3$\,dex depleted from the BBN value. The extent of AddMix is mainly due to the density, which falls off as $\rho^{-3}$ (using solar models while trying to represent Li destruction but avoiding Be destruction) and depends to a lesser extent also on the efficiency. The efficiency can be empirically constrained by the $A$(Li) value of the plateau and abundance trends of heavier elements. \\

Although a number of suggestions have been made towards the physical origin of AddMix (mass loss, rotation-induced mixing, etc.), the true underlying mechanism is still unknown. 
To further our knowledge about the intrinsic stellar processes involved, we turn to stars in metal-poor Galactic globular clusters. All stars in globular clusters are, to first approximation, born at the same time. This allows the evolutionary status of the observed stars to be determined unambiguously. And although the surface metallicities observed today may vary between stars in different evolutionary phases, due to the effects of atomic diffusion, e.g. \citet{Korn2007, Lind2008, Nordlander2012, Gruyters2013, Gruyters2014}, all stars were also born with the same (iron-peak) metallicity.\\

At a metallicity of \FeH\ $=-2.3$, M30 (NGC\,7099) is one of the most metal-poor globular clusters in the Milky Way. Its location far from the Galactic plane ensures a low reddening. However, given its large distance of $8.3\pm0.2$\,kpc \citep{Kains2013} and an absolute visual magnitude of $V=7.1$ \footnote{SIMBAD Astronomical Database (Centre de Donn\'ees astronomiques de Strasbourg)}, the globular cluster M30 is a faint target with typical turnoff point (TOP) stellar magnitudes of $V \sim 18.5$\,mag. This makes it very hard to obtain good quality spectra from turnoff stars and even more difficult to precisely analyse them. It is
also probably one of the reasons why, to our knowledge, this is the first attempt to study the chemical content of TOP stars in this cluster. Information on the chemical content of red giant branch (RGB) stars of M30 has been presented by \citet{Carretta2009a}. They detected a weak sodium-oxygen anticorrelation in their observed RGB stars. This anticorrelation is thought to be the direct result of the pollution of the star-forming gas by short-lived intermediate-mass or massive stars of the first generation of globular cluster stars \citep{Prantzos2006a,Decressin2007a,Dercole2010,Krause2013,Ventura2013,Denissenkov2014}. During their short lifetimes, these stars eject hydrogen-processed material and contaminate the intra-cluster gas from which a subsequent generation of stars is formed. Compared to the pristine cluster composition, the pollution mechanism will increase the Na abundances while lowering the O abundances in the cluster gas, making all second-generation long-lived low-mass stars seem chemically different from their first-generation counterparts. The observed anticorrelation in M30 thus presents evidence that M30 is harbouring at least two stellar generations. As Na also theoretically anticorrelates with Li, the same mechanism will  lower the Li abundance in second-generation stars as the more massive stars are hotter and thus burn the Li before it is returned to the interstellar medium. Thus, if we wish to obtain the primordial value of Li, we have to find a way to disentangle the effects of atomic diffusion and intrinsic stellar depletion from early cluster pollution \citep{Lind2009,Nordlander2012,Gruyters2014}.\\
In this work we present the results of a homogeneous abundance analysis of 144 stars in M30. We derived non-local thermal equilibrium
(NLTE) abundances for Li, Ca, and Fe in an attempt to trace the evolution of Li and derive new constrains on atomic diffusion and additional mixing. The outline of the paper is as follows. Section 2 describes the data sets, in Sect. 3 we present our analysis, and the results are given in Sect. 4. These results are discussed in Sect. 5, and a summary of the work can be found in Sect. 6.



\section{Observations}\label{sect:obs}
\subsection{Photometry and target selection}

\begin{figure}
\begin{center}
\includegraphics[width=1\columnwidth]{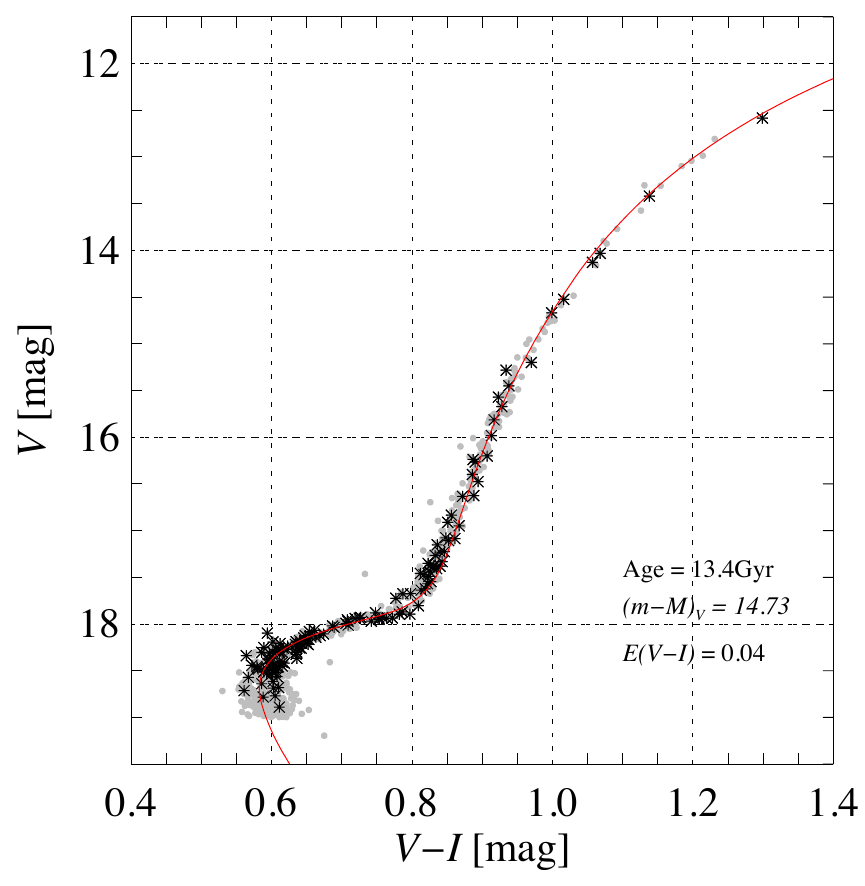}
\caption{Color-magnitude diagram of the globular cluster M30. The asterisks represent our target stars for which we obtained spectra. Overplotted is the Victoria isochrone at an age of 13.4\,Gyr and a metallicity of \FeH\ $= -2.30$.
}\label{Fig:CMD}
\end{center}
\end{figure}

The observations of stars in M30 are based on a carefully selected sample of 150 stars. The selection of stars was made on $uvby$  Str\"omgren photometry obtained with the Danish 1.54m telescope on La Silla, Chile \citep{Grundahl1999} and designed to obtain a homogenous sample covering all evolutionary phases and populations from the TOP to the tip of the RGB in the globular cluster (GC) M30. The focus of this study, however, lies upon the TOP as we aim to obtain a measurement of the Li abundance in M30. \\
No standard star observations were made during the observing run, therefore the photometry could not be properly calibrated and hence cannot be used to obtain precise stellar parameters. Instead, we used $VI$ broadband photometry from Peter Stetson \citep{Stetson2000,Stetson2005} to derive photometric stellar parameters. The photometry is displayed in Fig.\,\ref{Fig:CMD}, which shows the $V-I$ colour-magnitude diagram in which our targets are marked by asterisks.

\begin{figure}
\begin{center}
\includegraphics[width=1\columnwidth]{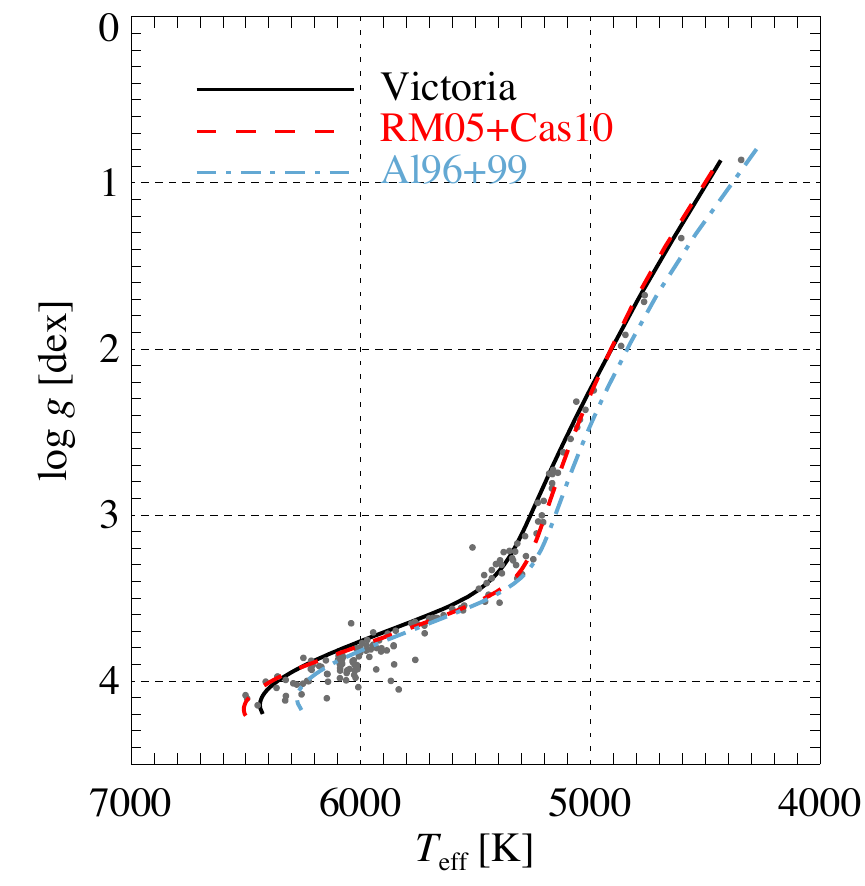}
\caption{Three effective temperature scales based on the $V-I$ photometric indices as a function of the $V$ magnitude. Overplotted as bullets are the individual effective temperatures.}
\label{Fig:fiducials}
\end{center}
\end{figure}

\subsection{Spectroscopy}
We collected 20 exposures of 45\,min centred on the globular cluster M30, using the FLAMES instrument on VLT-UT2, ESO Paranal \citep{Pasquini2003}. Approximately 100 fibres, connected to the medium-resolution spectrograph GIRAFFE, were allocated to stars in all evolutionary stages between the cluster TO point at $V=18.5$ and the tip of the RGB at $V=12.5$. As only the faintest targets required 20 exposures to obtain the required signal-to-noise
ratio, we used two different fibre configurations. This allowed us to switch fibres for the brighter stars. In total, we collected data for 150 stars using the setup HR15N H665.5 (R $\approx$ 20\,000). A minor fraction (four stars) of the data had to be discarded because of crosstalk between adjacent bright and faint objects on the CCD. Simultaneously, we collected high-resolution spectra with the UVES spectrograph for 13 RGB stars in total, using the standard setup centred on 580 nm. In this paper we focus on the GIRAFFE data set, the UVES data will be analysed and discussed in a subsequent paper.\\
The data were reduced with the dedicated ESO-maintained pipelines and further processed in IDL. For the majority of GIRAFFE targets, 20 individual exposures were co-added after sky-subtraction and radial-velocity correction. In this process, one foreground star (id 200161) with radial velocity $-65$\,\kms was identified, as well as clear signs of binarity for one SGB star (id 62198). These stars were excluded from the analysis. The mean heliocentric radial velocity of the cluster was found to be $-187$\,\kms, with a dispersion of $\sigma = 3$\,\kms and agrees well with the values given by \citet{Carretta2009a} ($-188\pm5$\,\kms) and \citet{Harris1996} ($-184.2\pm0.2$\,\kms, latest web update (2010)).



\section{Analysis}\label{sect:analysis}
\subsection{Effective temperature}
The effective temperatures for the target stars were derived by fitting Victoria isochrones \citep{Vandenberg2014} transposed to the $V$--$(V-I)$ colour-magnitude space using the latest synthetic colour transformations of \citet{Casagrande2014} (see Fig.\,\ref{Fig:CMD}). The Victoria isochrones are similar to the Montreal-Montpellier isochrones \citep[for a comparison see e.g.][]{Vandenberg2002,Vandenberg2012}, the only difference being that the models do not include radiative acceleration of elements or additional mixing. The best fit to the photometry is given by an isochrone with an age of 13.4\,Gyr and a reddening of $E(V-I) = 0.037$. The distance modulus corresponding to this best fit is 14.73 and falls in between 14.64 and 14.82, the values given in the Harris catalogue \citep[][latest web update (2010)]{Harris1996} and by \citet{Dotter2010}, respectively. Even though we also tried BaSTI \citep{Pietrinferni2013} and Dartmouth isochrones \citep{Dotter2008}, the Victoria isochrones are the only isochrones that can reproduce the morphology of the cluster within a reasonable parameter space (age <15Gyr, \Teff(TOP)$<6600$\,K). \\

The effective temperatures were then obtained by projecting the stars onto the isochrone along the line of shortest distance perpendicular to the isochrone. This ensures that the shift in colour will not become unphysically large on the flat part of the colour-magnitude diagram (CMD),  the SGB. We compared the obtained temperature scale with two different calibrations based on published relations between \Teff\ and $V-I$ colour indices, both calibrated on the infra-red flux method \citep[IRFM]{Blackwell1986}. A first relation we adopted from \citet{Alonso1996} and \citet[][hereafter these two papers are referred to as Al96+99]{Alonso1999}.

As a second calibration we combined the calibrations by \citet[][Cas10 hereafter]{Casagrande2010} for the TO and SGB stars, with the calibration by \citet[][RM05 hereafter]{Ramirez2005} for the RGB stars. A zero-point correction of +100 K was applied to the RM05 scale to bring RM05 to the same scale as that of Cas10 and allow for a smooth connection to the colour-\Teff\ relation of the less evolved stars. For all stars we assumed \FeH$=-2.3$ when applying the relations. As IRFM colour-\Teff\ relations are constructed for either dwarfs ($\log g \geqslant 3.8$) or giants ($\log g \leqslant 3.5$), we interpolated linearly between each dwarf- and giant-calibration pair to obtain a smooth calibration for the SGB. The interpolation was made for a $V-I$ range between 0.62 and 0.8 mag.\\
Figure\,\ref{Fig:fiducials} shows the three $V-I$-based \Teff\ scales as a function of $\log g$. While the Victoria and Alonso effective temperature scales are offset by 150\,K on average, the difference with the RM05+Cas10 \Teff\ scale is more variable, as the \Teff\ scale has a different morphology than the others. While the RM05+Cas10 \Teff\ scale (dashed line) is about 70\,K hotter than the Victoria \Teff\ scale at the TOP (\Teff $\geqslant6300$), it is on average about 70\,K cooler on the base-RGB (bRGB) before converging to the Victoria scale on the RGB. On average, however, the RM05+Cas10 scale is 17\,K cooler than the Victoria scale.\\

Additionally, a spectroscopic \Teff\ scale was also derived by fitting the wings of H$\alpha$ using a grid of 1D LTE plane-parallel and spherical MARCS model atmospheres \citep{Gustafsson2008} together with the spectral synthesis code Spectroscopy Made Easy \citep[SME,][]{Valenti1996,Valenti2005}. The automated fitting method uses metal-line free regions extending up to $\pm50$\,\AA\ from the centre of the H$\alpha$ line \citep[see][]{Lind2008} while avoiding the line core. We note that this method is not
applicable to the coolest stars in the sample because temperature sensitivity of the wings of the Balmer lines becomes too weak. For those stars we used lines from iron-peak elements (see Table\,\ref{Tab:lines}) to obtain the \Teff\ from the excitation balance. {\bf As the spectra of the hottest stars are the result of a coaddition of up to 20 exposures
that were stacked with an outlier-resistant technique, they do not show strong contamination of tellurics. If weak tellurics are present, they do not strongly influence the best-fit temperature since they are evenly distributed over the H$\alpha$ profile and neighbouring continuum regions. Consequently, the uncertainty is set by the signal-to-noise ratio in the line.}

\begin{figure}
\begin{center}
\includegraphics[width=1\columnwidth]{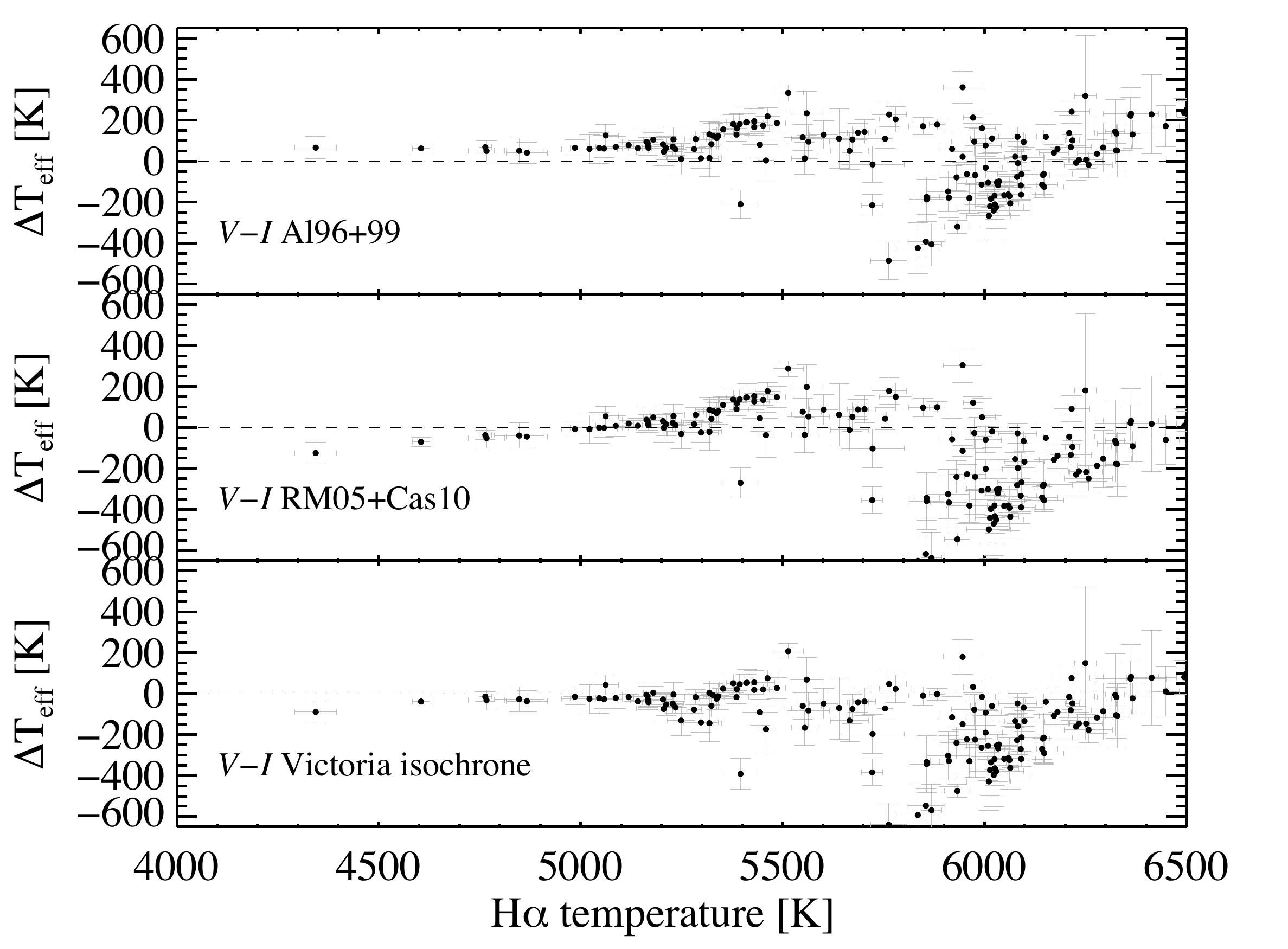}
\caption{Comparison between the H$\alpha$-based \Teff\ scale and the photometric \Teff\ scales obtained by applying the calibrated relations of \citet{Alonso1996,Alonso1999} and \citet{Casagrande2010}+\citet{Ramirez2005}, to colour indices $V-I$. The y-axis shows $\Delta$\Teff\ = \Teff(H$\alpha$) $-$ \Teff(photometry) for the calibrations. For stars with temperatures below 4950\,K, the excitation temperatures are shown. The corresponding error bars are overplotted in grey. Horizontal error bars correspond to the fitting uncertainty, vertical error bars represent the uncertainty in the photometric temperatures as given by the spread in colour around the fiducial, added in quadrature with the uncertainty in the spectroscopic temperatures as given by the fitting error.
}\label{Fig:Teff_comparison}
\end{center}
\end{figure}

A comparison between the photometric \Teff\ scales and the spectroscopic one is given in Fig.\,\ref{Fig:Teff_comparison}. The plot shows the difference between the derived H$\alpha$ temperatures and the effective temperatures obtained from the different photometric calibrations and the isochrone temperatures.From the comparison we see that the Victoria \Teff\ scale agrees best with the H$\alpha$ temperatures. This is especially true on the bRGB (5000\,K$\leqslant$ \Teff $\leqslant5500$\,K), where the Victoria scale is on average $30\pm80$\,K hotter than the Al96+99 and RM05+Cas10 scales, which are $97\pm73$\,K cooler and $49\pm77$\,K cooler, respectively, than the H$\alpha$ temperatures. On the TOP (\Teff $\geqslant6300$) the Victoria scale also matches the H$\alpha$ temperatures best with an average difference of $17\pm78$\,K compared to $172\pm76$\,K and $-48\pm83$\,K for the Al96+99 and RM05+Cas10 scales, respectively. On the SGB the temperatures seem less well constrained, as is apparent from the large spread in H$\alpha$ temperatures (see Fig.\,\ref{Fig:fiducials}). This is largely due to the decrease in the signal-to-noise ratio with increasing $V$ magnitude, but also shows how difficult it is to achieve high precision on the temperatures of SGB stars using a photometric calibration. \\

Given the high uncertainty on \Teff\ based on H$\alpha$ compared to photometric \Teff\, and the large difference between the Al96+99 and RM05+Cas10 \Teff\ scale, we here opt to use temperatures based on the isochrone as our primary \Teff\ scale because it is the most homogeneous. We note that we achieve the better agreement with the Victoria scale partly by construction, as we varied age, distance modulus and reddening to obtain the best fit with the observations. Circular arguments with respect to atomic diffusion are not an issue. The complete set of effective temperatures is given in Table\,\ref{Tab:SP}.

\begin{table*}
\caption{Photometry and effective temperatures. The full table can be retrieved from CDS/Vizier.}
\label{Tab:SP}
\centering
\begin{tabular}{rcc cc cccc }
\hline\hline
 ID &  RA (J2000) & Dec (J2000) & $V$     & $V-I$     & \Teff\,Vic  & \Teff\,RM05+Cas10 & \Teff\,Al96+99 & \Teff\,H$_{\alpha}$ \\
     &                      &                      &            &               &   [K] &   [K] &   [K] &   [K]  \\
 \hline 
36 & 21 40 16.85 & -23 5 10.40 & 18.438 & 0.593 & 6379 & 6447 & 6226 & 6294 \\
44 & 21 40 21.12 & -23 5 15.50 & 18.037 & 0.692 & 5938 & 5850 & 5758 & 5972 \\
49 & 21 40 13.98 & -23 5 23.30 & 18.580 & 0.584 & 6425 & 6499 & 6269 & 6064 \\
98 & 21 40 11.86 & -23 6  3.60 & 18.723 & 0.583 & 6439 & 6508 & 6276 & 6011 \\
101 & 21 40 27.53 & -23 6  1.90 & 18.460 & 0.591 & 6389 & 6459 & 6236 & 6026 \\
108 & 21 40 19.16 & -23 6  7.20 & 18.268 & 0.618 & 6263 & 6310 & 6114 & 6009 \\
$\cdots$ & $\cdots$ &  $\cdots$ &$\cdots$&$\cdots$&$\cdots$&$\cdots$ & $\cdots$ & $\cdots$ \\
\hline
\end{tabular}
\end{table*}

\subsection{Surface gravities}
As we use temperatures based on the isochrone, we complemented these temperatures with the corresponding gravities to complete our stellar parameter set homogeneously. Using the 13.4\,Gyr isochrone, we find surface gravities ranging from $\log g =$ 4.21 on the TOP at 6400\,K to $\log g =$ 0.82 at the tip of the RGB at 4200\,K. 

To estimate the uncertainty on the gravities, we investigated the difference between the adopted isochrone gravities and photometric gravities. Photometric surface gravities are commonly derived by using the customary relation between the effective temperature, the luminosity, the mass, and the surface gravity. The luminosities were calculated using the bolometric correction that is a function of metallicity \FeH\ and \Teff\ and were obtained by using the \citet{Alonso1999} calibration. Given the two photometric \Teff\ scales, we can calculate the difference in gravity for each scale. We find very similar gravities, with average differences of $-0.033$ and $+0.008$ for the Al96+99 and RM05+Cas10 scales, respectively.

The study presented here is based on the abundance difference between stars. As such, the importance lies in the accuracy in surface gravities on a relative scale rather than on the absolute scale. The dominant source of errors then stems from the uncertainty in \Teff, for example an error in \Teff\ of +100\,K translates into an increase of approximately 0.03 dex in logarithmic surface gravity. This is why we find a larger difference between the Victoria gravities and the gravities derived by using the RM05+Cas10 \Teff\ scale than for the Al96+99  \Teff\ scale.
Other contributions to the error on $\log g$ are minor compared to the uncertainty stemming from \Teff. An uncertainty in stellar mass of 0.03\,M$_{\odot}$ leads only to an uncertainty of 0.015\,dex, while an increase of 0.01 mag in $V$ results in an error in $\log\,g$ of about 0.004\,dex. Finally, the uncertainty in distance modulus will only affect the absolute values of the surface gravity, for example an increase of 0.5 in $(m-M)_V$ will raise the overall $\log g$ with 0.2\,dex. \\
The expected precision in $\log g$ thus predominantly follows the uncertainty in \Teff\ , and given the colour-based uncertainty of 30\,K in the giants and 85\,K in the dwarfs, the surface gravity is accurate within 0.02\,dex for the giants and 0.03\,dex for the dwarfs. We note that the uncertainty on distance modulus is not included {\bf and that, using the uncertainties based on \Teff(H$\alpha$), we find uncertainties in the gravity of 0.1 dex on the TOP (\Teff\ $>5900$\,K with a \Teff\ uncertainty of 161\,K) and 0.03 dex on the RGB (\Teff\ $<5200$\,K with a \Teff\ uncertainty of 50\,K).}

\begin{table}
\caption{Selection of lines used to determine temperature and abundance.}
\label{Tab:lines}
\centering
\begin{tabular}{lcrrcc}\hline\hline
Ion & Wavelength$[\AA]$ & $E_{\mbox{{\scriptsize low}}}$[eV] & $\log gf$ & TOP & RGB \\ 
 \hline
H\,{\scriptsize I} & 6562.7970 & 10.199 & 0.710 & X & X \\
Li\,{\scriptsize I} & 6707.7635 & 0.000 & -0.002  & X & X \\
Li\,{\scriptsize I} & 6707.9145 & 0.000 & -0.303 & X & X \\
Ca\,{\scriptsize I} & 6493.7810 & 2.521 & -0.109 & X & X \\
Ca\,{\scriptsize I} & 6499.6500 & 2.523 & -0.818 & X & X \\
Ca\,{\scriptsize I} & 6717.6810 & 2.709 & -0.524 & X & X \\
Ti\,{\scriptsize I} & 6743.1221 & 0.900 & -1.611 &   & X \\
Ti\,{\scriptsize II} & 6491.5659 & 2.061 & -1.942 &   & X \\
Fe\,{\scriptsize I} & 6494.9804 & 2.404 & -1.268 & X & X \\
Fe\,{\scriptsize I} & 6498.9383 & 0.958 & -4.687 & X & X \\
Fe\,{\scriptsize I} & 6518.3657 & 2.832 & -2.460 & X & X \\
Fe\,{\scriptsize I} & 6591.3128 & 4.593 & -2.081 &   & X \\
Fe\,{\scriptsize I} & 6593.8695 & 2.433 & -2.420 &   & X \\
Fe\,{\scriptsize I} & 6663.4411 & 2.424 & -2.479 & X & X \\
Fe\,{\scriptsize I} & 6677.9851 & 2.692 & -1.418 & X & X \\
Fe\,{\scriptsize II} & 6516.0766 & 2.891 & -3.310 & X & X \\
Ni\,{\scriptsize I} & 6586.3098 & 1.951 & -2.746 &    & X \\
Ni\,{\scriptsize I} & 6643.6303 & 1.676 & -2.300 &  & X \\
Ni\,{\scriptsize I} & 6767.7720 & 1.826 & -2.170 &  & X \\
Ba\,{\scriptsize II} & 6496.8970 & 0.604 & -0.407 & X & X \\
Eu\,{\scriptsize II} & 6645.0940 & 1.380 & 0.120 &  & X \\
\hline
\end{tabular}
\tablefoot{Only the lines measurable in the TOP stars were used to determine the abundance, the other lines were only included to derive a spectroscopic temperature for the RGB stars.}
\end{table}

\subsection{Spectral synthesis}
We determined spectroscopic temperatures (to compare to our photometric \Teff\ values) and chemical abundances for individual elements with an automated pipeline,  based upon the SME spectrum synthesis program \citep{Valenti1996}. The pipeline consists of a suite of IDL routines, optimised to control the analysis of Gaia-ESO GIRAFFE data through an iterative scheme. Briefly, we performed a $\chi^{2}$ minimisation between the observed spectra and synthetic spectra based upon a grid of 1D LTE plane-parallel and spherical MARCS model atmospheres \citep{Gustafsson2008}, applying masks to un-blended parts of carefully selected spectral lines (see Table\,\ref{Tab:lines}). The observed spectra ingested in the pipeline are not pre-normalised.

The iteration scheme used in the pipeline is divided into two main blocks. For the first block, we adopted the starting guesses for \Teff, and $\log g$ based on photometry, and kept \FeH\ fixed to the metallicity of the cluster with an $\alpha$-enhancement of 0.4\,dex. The second block adopted as starting guess the final results of the first block.  Each block was divided into two parts, normalisation and stellar-parameter determination. 

The normalisation was performed before each stellar parameter determination run by robust linear fits to 5-60~\AA\ segments, minimising the $\chi^{2}$  distance between observations and a fixed-parameter synthesis. Thereafter, the $\chi^{2}$ minimisation was designed to optimise the \Teff. These derived \Teff\ values were then used as our spectroscopic \Teff\ scale.

The micro- and macro-turbulence values were determined iteratively, using the starting parameter values for each block, from the relations used in the Gaia-ESO survey. The projected rotational velocity was set to 1.0~\kms.

The main physical diagnostics that governs the stellar parameter optimisation were the H$\alpha$ line for \Teff\ and, if possible, the excitation balance of Fe~I lines. In each iteration the $\log g$ was updated by using the customary relation between the effective temperature, the luminosity, the mass, and the surface gravity.\\

In a second iteration block, we solved for individual chemical abundances by using as input stellar parameters the different sets, spectroscopic, photometric, and isochrone-based stellar parameters. 
 To this aim, masks covering unblended spectral features for Li, Ca, and Fe,  (Li, Ca, Ti, Fe, Ni, and Ba) were automatically computed for each star (RGB star)  and adopted to determine the abundance if the number of pixels in the mask exceeded five. Additionally, the code was modified to allow NLTE line formation for Li, Fe, and Ca from a grid of precomputed corrections (tabulations of LTE departure coefficients). The NLTE corrections for Li and Fe are based on pre-computed departure coefficients by \citet{Lind2009} and \citet{Lind2012} for Li and Fe, respectively. The Ca NLTE corrections are based upon the Ca NLTE model atom described in footnote 3 on page 7 of \citet{Melendez}.

The errors provided by SME are computed from the error vectors constructed from the variance of the observations. These form part of the definition of the $\chi^{2}$ in SME and are thus used to optimise parameters, derive abundances, and determine their associated internal errors. The given uncertainties in stellar parameters have not been propagated and added into these abundance uncertainties.


\section{Results}
\subsection{Ca and Fe}
Iron and calcium abundances were derived for a sample of 144 stars.  
As the stars were only observed in one GIRAFFE setting (HR15N), the line selection is very limited. 
For Ca, the spectrum shows one `strong' (EW=15-20\,m\AA\ in TOP spectrum) Ca line at 6493.9\,\AA\, and two weak to very-weak lines (EW=5-10\,m\AA\ in TOP spectrum) at 6499.7 and 6717.7\,\AA. For Fe, we find one strong line at 6495.0\,\AA\ and five very weak lines, including one Fe\,{\scriptsize II} line at 6516.1\,\AA. 

\begin{table*}
\caption{Average abundances based on the individual spectra based in turn on the different temperature scales.}
\label{Tab:individ-trends}
\centering
\begin{tabular}{lcccrr}\hline\hline
\Teff-scale & $\Delta T_{\mbox{\scriptsize eff}}$ & $\Delta \log g$ & $\Delta \xi$ & \ch{$\Delta \abund{Ca}$\tablefootmark{a}}  & \ch{$\Delta \abund{Fe}$\tablefootmark{b}}  \\ 
              & (K)     & (cgs) & (km\,s$^{-1}$) & \ch{NLTE} & \ch{NLTE}  \\ \hline
Victoria           & $1282\pm86$ & $1.59\pm0.13$ & $0.34\pm0.06$ & $0.05\pm0.71$ & $-0.33\pm0.35$ \\
Al96+99          & $1299\pm106$ & $1.60\pm0.21$ & $0.32\pm0.06$ & $0.07\pm0.39$ & $-0.37\pm0.39$ \\
RM05+Cas10 & $1375\pm135$ & $1.60\pm0.22$ & $0.39\pm0.09$ & $0.03\pm0.60$ & $-0.39\pm0.43$ \\
H$\alpha$ & $1259\pm128$ & $1.61\pm0.18$ & $0.30\pm0.08$ & $0.05\pm0.27$ & $-0.43\pm0.33$ \\
\hline
\end{tabular}
\tablefoot{The differences were calculated based on the mean values derived from individual stars on the TOP (\Teff $>6200$\,K) and the middle of the RGB ($5100$\,K $<$ \Teff $<4900$\,K) and include the 1-$\sigma$ uncertainties.
\tablefoottext{a}{Based on three Ca\,{\scriptsize I} $\lambda\lambda$6493.9, 6499.7 and 6717.7.}
\tablefoottext{b}{Based on five Fe\,{\scriptsize I} and one Fe\,{\scriptsize II} lines $\lambda\lambda$6495.0, 6518.4, 6592.9,  6593.9, and 6678, and $\lambda$6516.1.}
}
\end{table*}

\begin{figure}
\begin{center}
\includegraphics[width=1\columnwidth]{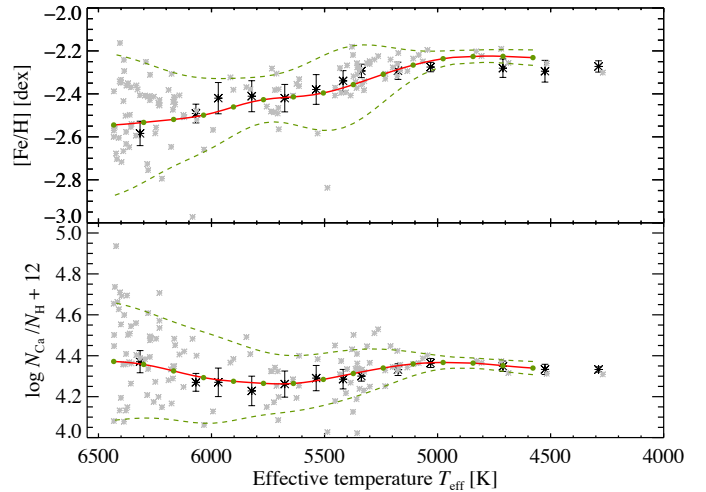}
\caption{Derived NLTE abundances as a function of the effective temperature \Teff\ for Fe and Ca from the analysis of 144 stars. The solid (red) lines represent the running mean (weighted average), the dashed (green) lines give the standard deviation. Overplotted are the derived abundances from the coadded group-spectra with their respective statistical errors. 
}\label{Fig:individ_abund}
\end{center}
\end{figure}

The derived NLTE abundances for iron and calcium are displayed in Fig.\,\ref{Fig:individ_abund}. The measurements become more uncertain with increasing \Teff\ as the lines become weaker and the signal-to-noise ratio drops. This is shown in the figure by the increasing standard deviation (dashed lines) around a running mean (solid line). The running mean shows the weighted average abundance in bins of $\pm150$\,K, smoothed by a Gaussian ($\sigma = 100$\,K). To increase our chances of detecting Fe and Ca abundances near the hot end of the sample, we coadded stars with similar stellar parameters, weighting each spectrum with respect to its signal-to-noise ratio, as we did in previous papers in the series. The abundances derived from these coadded group spectra are overplotted in Fig.\,\ref{Fig:individ_abund} with their respective error bars that represent the statistical error. We omit the error bars for the individual stars for reasons of clarity. \\
Fe shows a trend with effective temperature, while Ca seems more or less flat. We note that the coadded group-spectra display the same abundance pattern as the running mean of the individual stars. Similar (slightly stronger) trends are obtained with the other \Teff-scales discussed in the paper. The different trends are given in Table\,\ref{Tab:individ-trends}, while Table\,\ref{Tab:coadd-trends} summarises the average abundances derived for Fe and Ca at three representative effective temperature points corresponding to the RGB, SGB, and TOP. The values in Table\,\ref{Tab:individ-trends}   are derived by taking the difference between the mean values and their corresponding standard deviations on the TOP (\Teff $>6200$\,K) and the middle of the RGB ($5100$\,K $<$ \Teff $<4900$\,K). In Table\,\ref{Tab:coadd-trends} the values are derived by taking the median values within each evolutionary group. The uncertainties on the abundance are given by the corresponding standard deviation within each group. The results for the individual coadded groups can be found in Table\,\ref{Tab:coadd-results} in the appendix. 

\begin{table*}
\caption{Average abundances based on the coadded spectra and obtained at three effective temperature points.}
\label{Tab:coadd-trends}
\centering
\begin{tabular}{lcccrr}\hline\hline
Group & $T_{\mbox{\scriptsize eff}}$ & $\log g$ & $\xi$ & \ch{$\abund{Ca}$\tablefootmark{b}}  & \ch{$\abund{Fe}$\tablefootmark{d}}  \\ 
              & (K)     & (cgs) & (km\,s$^{-1}$) & \ch{NLTE} & \ch{NLTE}  \\ \hline
TOP       & 6327 & 3.98 & 1.46 & $4.39\pm0.61$ & $4.98\pm0.30$  \\
SGB       & 5676 & 3.55 & 1.12 & $4.32\pm0.14$ & $5.08\pm0.19$  \\
RGB       & 4718 & 1.69 & 1.27 & $4.37\pm0.03$ & $5.23\pm0.04$  \\
\hline
$\Delta(\text{TOP}-\text{RGB})$  & 1609 & 2.29 & 0.19 & $+0.02$ & $-0.25$  \\
 \hline
\end{tabular}
\tablefoot{The average stellar parameters for the TOP/SGB/RGB stars are based on the averages from the coadded spectra with the warmest or coolest three \Teff\  values.
\tablefoottext{b}{Based on three Ca\,{\scriptsize I} $\lambda\lambda$6493.9, 6499.7 and 6717.7.}
\tablefoottext{d}{Based on five Fe\,{\scriptsize I} and one Fe\,{\scriptsize II} lines $\lambda\lambda$6495.0, 6518.4, 6592.9,  6593.9, and 6678, and $\lambda$6516.1.}
}
\end{table*}

\subsection{Li}
In addition to Ca and Fe abundances, we also derived lithium abundances by using the $^7$Li resonance line at 6707\AA. The line has two fine-structure components, separated by merely 0.15\,\AA\ and hence unresolved at the GIRAFFE resolution of $R=19\,300$. To measure the line accurately, we used atomic-line data from the Vienna Atomic Line Database \citep[VALD,][]{Piskunov1995,Kupka1999}. We adopted $\log gf = -0.036$ and $-0.337$  for the different components.\\
Li abundances where derived under the assumption of LTE and corrected for NLTE effects in 1D according to \citet{Lind2009a}. The corrections on Li at this metallicity are rather small with typical values of $-0.05$\,dex for TOP and SGB stars, but they increase and change sign with decreasing \Teff\ and $\log g$. The largest corrections ($+0.13$ - $+0.15$\,dex) are found for the coolest RGB stars. The derived NLTE abundances can be found in Table\,\ref{Tab:individ Abund} along with the adopted stellar parameters and the abundances for Ca and Fe. Figure\,\ref{Fig:Li diffusion} shows the derived NLTE Li abundances as a function of absolute visual magnitude $M_V$ along with the CMD of M30. The figure agrees with expectations: we find two very well defined plateaus, one for dwarfs ($A$(Li) $\approx 2.2$) and one for giants ($A$(Li) $\approx 1.1$). At the end of the dwarf plateau in the middle of the SGB ($M_V\approx3.25$), the Li abundances drastically drop to the giant plateau as a result of the first dredge-up, the dilution of the stellar surface convection layer with hydrogen-processed and lithium-depleted material from deeper layers. The giant plateau ends at the RGB bump ($M_V \approx 0$) with another steep drop in Li abundance. This last drop can probably be explained by thermohaline mixing \citep{Charbonnel2007}, a mixing process that becomes efficient when the hydrogen-burning shell crosses the chemical discontinuity left behind by the first dredge-up. This mixing process rapidly transports the surface Li down to the hotter inner layer where the fragile element is readily destroyed. The result is a Li-depleted surface layer. Similar lithium trends are observed in other metal-poor clusters such as NGC\,6397 \citep{Lind2009,Nordlander2012}, NGC\,6752 \citep{Gruyters2013,Gruyters2014}, and M4 \citep[][Gruyters et al., submitted]{Mucciarelli2012}, \citep{GruytersM4} as well as in halo field stars \citep{Gratton2000}. \\

\begin{figure*}
\begin{center}
\includegraphics[width=0.95\columnwidth]{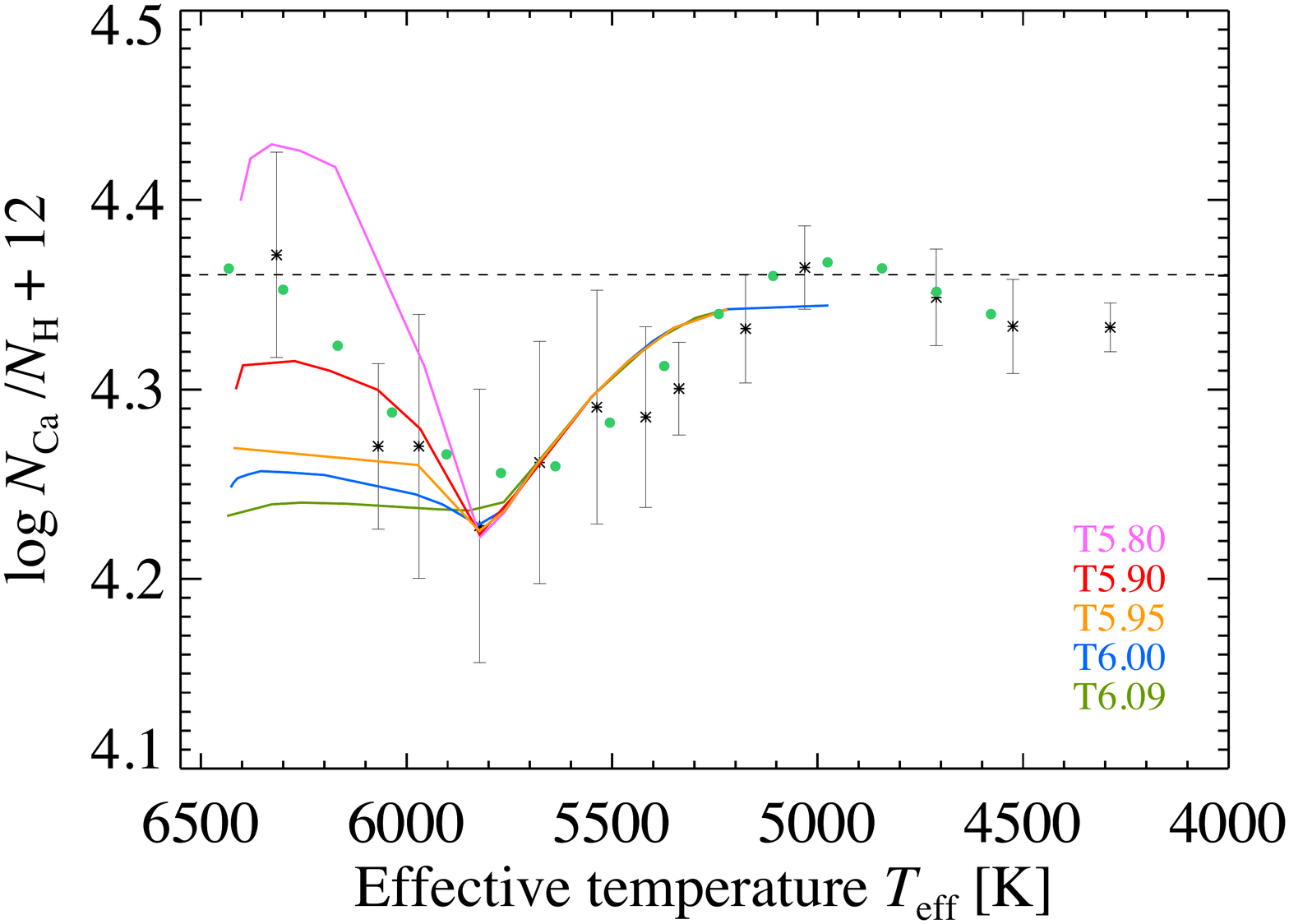}
\includegraphics[width=0.95\columnwidth]{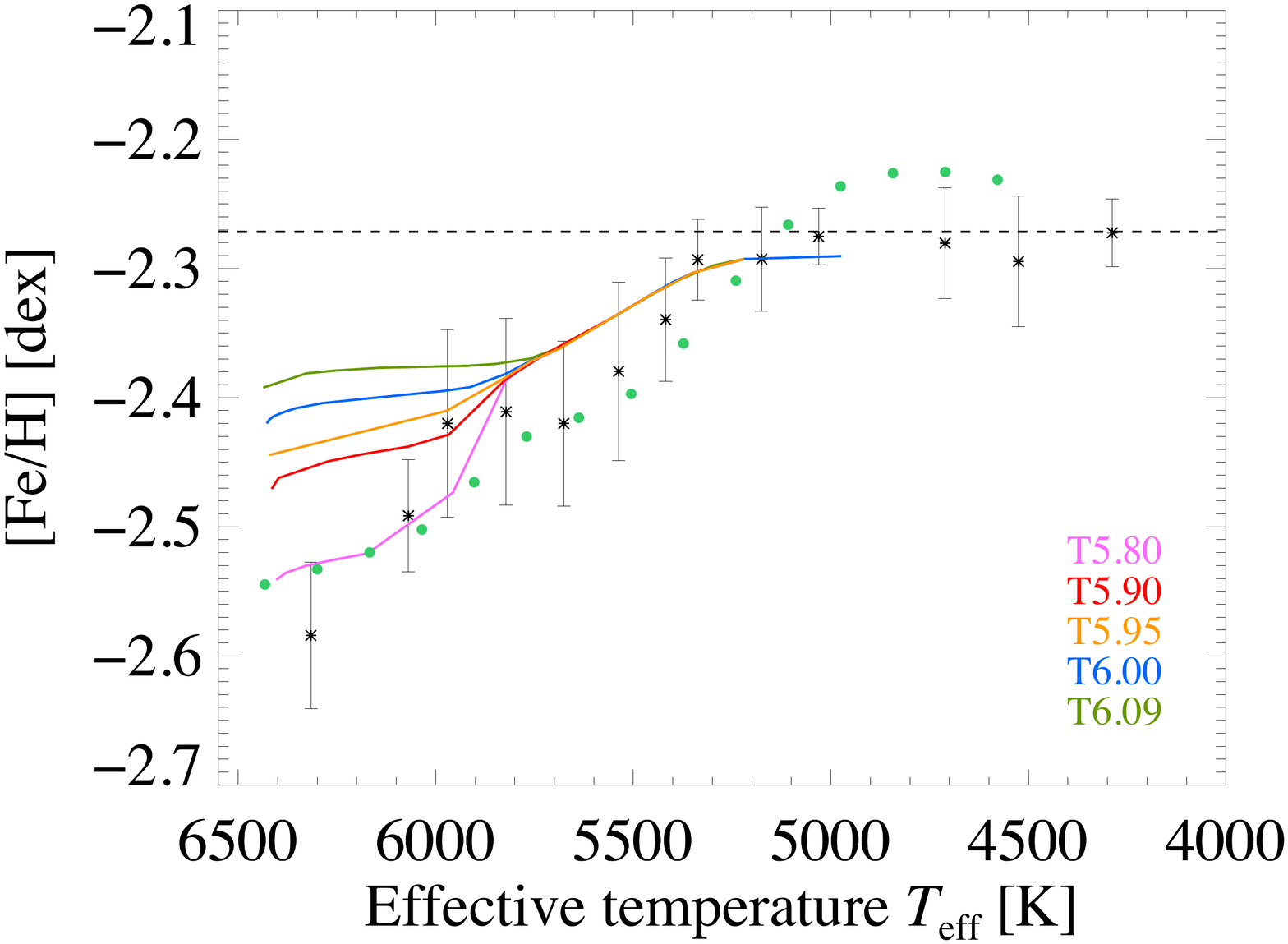}
\caption{Average abundance trends from Fig.\,\ref{Fig:individ_abund} given by the (green) bullets with the abundances of the coadded group-spectra overplotted. In both panels, predictions from stellar structure models including atomic diffusion with additional mixing with different efficiencies are overplotted. The dashed horizontal lines represent the initial abundances of the models, which have been adjusted so that predictions match the observed abundance level at the cool end of the \Teff\ scale.}
\label{Fig:diffusion trends}
\end{center}
\end{figure*}

\begin{table*}
\caption{Adopted stellar parameters and abundances of Li, Ca, and Fe.}
\label{Tab:individ Abund}
\centering
\begin{tabular}{rcc cc cccc}
\hline\hline
 ID &  \Teff & $\log g$ & $\abund{Li}$ & eLi   & $\abund{Ca}$  & eCa & $\abund{Fe}$ & eFe \\
     &    [K]  &  [dex]     &       NLTE       & [dex] &       NLTE       &  [dex]  &       NLTE       & [dex] \\
 \hline 
          36 &         6379 &  4.02 &  2.37 &  0.08 &  4.69 &  0.07 &  5.09 &  0.11  \\
          44 &         5938 &  3.73 &  2.30 &  0.07 &  4.37 &  0.07 &  5.13 &  0.08  \\
          49 &         6425 &  4.09 &  1.75 &  0.15 &  4.66 &  0.09 &  3.47 &  0.12  \\
         101 &         6389 &  4.04 &  2.19 &  0.10 &  3.60 &  0.17 &  5.09 &  0.11  \\
         108 &         6263 & 3.92  &  2.38 &  0.11 &  4.20 &  0.12 &  5.01 &  0.20  \\
$\cdots$ & $\cdots$ &  $\cdots$ &$\cdots$&$\cdots$&$\cdots$&$\cdots$ & $\cdots$ & $\cdots$ \\
\hline
\end{tabular}
\end{table*}

\section{Discussion}
\begin{figure}
\begin{center}
\includegraphics[width=1\columnwidth]{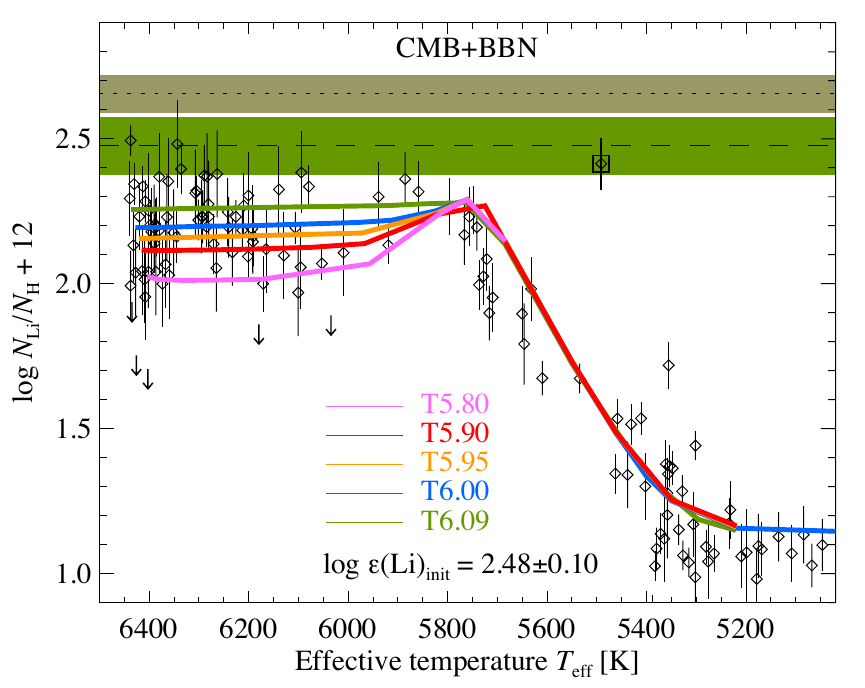}
\caption{Evolution of lithium compared to the prediction by Big Bang Nucleosynthesis (BBNS) given by the dotted line. The diamonds represent the observed Li abundances with their corresponding measurement error bars. The arrows represent upper limits. The square shows the Li-rich subgiant. Overplotted are predictions from stellar structure models including atomic diffusion and different efficiencies of AddMix. The initial abundance and corresponding uncertainty on the derived Li abundance are given by the dashed line and the shaded region, respectively. They have been adjusted so that predictions match the observed abundance level at the cool end of the \Teff\ scale. The shaded region around the dotted line gives the uncertainty interval of the PLANCK-calibrated primordial lithium abundance.
}\label{Fig:Li cosm}
\end{center}
\end{figure}

\subsection{Comparison to atomic diffusion models}
Figure\,\ref{Fig:diffusion trends} shows average abundances and the abundances from the coadded group-averages of Ca and Fe as a function of effective temperature. In both panels the observed abundance trends are compared with predictions from stellar-structure models including atomic diffusion and additional mixing (AddMix) with different efficiencies. While atomic diffusion is modelled from first principles, AddMix is a mixing mechanism modelled as a diffusive process using a parametric function of density. The efficiency of AddMix is given by a reference temperature $T_0$  \citep[][and references therein]{Richard2005}.  The absolute abundance scale of the models is slightly shifted to agree with the observations from the coadded group-averages for stars evolved beyond the onset of the first dredge-up. The dashed line gives the predicted initial abundance after correcting for atomic diffusion and AddMix. \\
Figure\,\ref{Fig:Li cosm} gives the evolution of Li as a function of effective temperature for the individual stars. In the plot, the arrows represent upper limits to the Li abundance, while the solid lines are the predictions from stellar-structure models including atomic diffusion and additional mixing (AddMix) with different efficiencies. For Li the efficiency of AddMix determines the amount of Li that is transported from the surface convection zone to the stellar interior.  
The efficiency thus affects the overall appearance of the Li abundance plateau. Where the T6.09 model (highest efficiency of AddMix in this paper) predicts a flat plateau, the T5.80 model clearly predicts the surface abundances to be dependent on effective temperature. Unfortunately, none of the models perfectly reproduces the derived Li abundances together with the observed abundances of Ca and Fe. The low Fe and high Ca abundances observed in the hottest coadded group-average, however, seem to point towards a model with a lower efficiency (T5.80), lower than that needed to explain the trends in NGC\,6397 at a metallicity of \FeH$=-2.1$. We do, however, advise caution about this result. As the size of the abundance trends predicted by the models is set by the relative abundance difference between the hottest and the coolest stars, it is crucial to derive precise abundances in these points. Given the limited signal-to-noise ratio and the weakness of the lines on the hot end of the sample, it is extremely difficult to reach a precision better than 0.1\,dex, which is needed to constrain the efficiency of AddMix. To obtain our warmest coadded group-average abundances, we coadded 28 spectra to reach a signal-to-noise ratio of about 60. This is just barely enough to derive Ca and Fe abundances with a measurement error of 0.1\,dex, which does not include propagated uncertainties based on stellar parameters. The true uncertainty on abundances thus does not allow us to strictly limit the efficiency of AddMix.\\

\begin{figure*}
\begin{center}
\includegraphics[width=1.75\columnwidth]{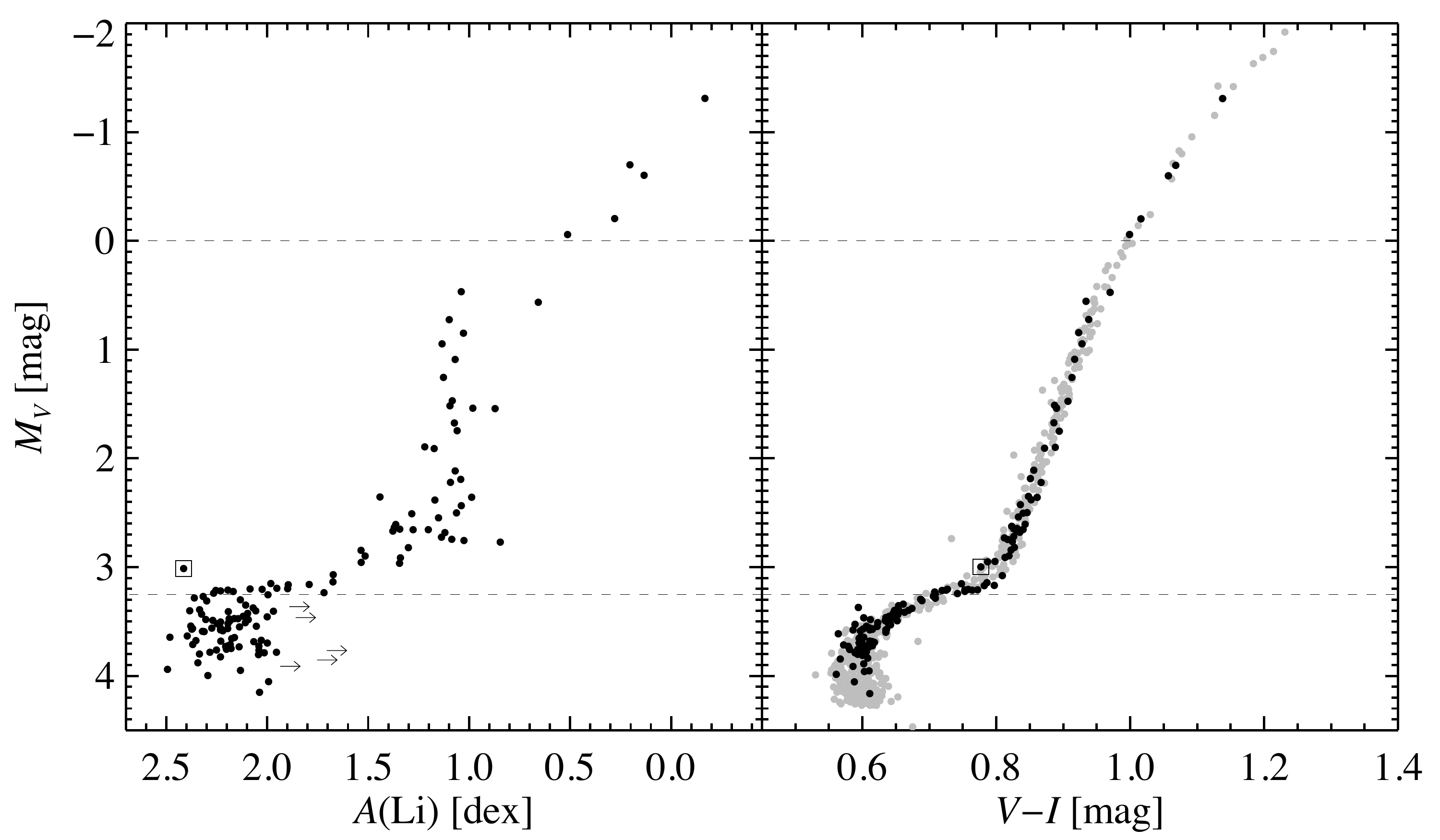}
\caption{\emph{Left:} Derived NLTE Li abundances as a function of the ordinate absolute visual magnitude ($M_V = V - 14.73$). \emph{Right:} CMD of M30. Our spectroscopic targets are marked by the black circles. The two horizontal dashed lines mark the locations where the Li abundance rapidly decreases as a result of stellar evolution. The square marks the Li-rich subgiant in the two plots (see text).
}\label{Fig:Li diffusion}
\end{center}
\end{figure*}

The initial Li abundance of M30 was obtained by matching the models to the observed Li abundances (see Fig.\,\ref{Fig:Li cosm}). As the observations do not allow us to place firm constrains on the efficiency of AddMix, we chose the T6.00 model to derive a conservative lower estimate of the initial Li abundance. We chose the T6.00 since this efficiency was derived for the GC NGC\,6397, which has a metallicity (\FeH\ $=-2.1$) close to that of M30, and we do not observe a bump in the Li abundances around 5800\,K that is predicted by the models with lower efficiency. After correcting for atomic diffusion and AddMix with an efficiency of T6.00, we find a value of $A(\text{Li})=2.48\pm0.10$. However, based on the observed abundances for Ca and Fe, it might also
be argued that the efficiency needs to be low. For the T5.80 model the initial Li abundance would then increase to $A(\text{Li})=2.68\pm0.10$ (the models would be shifted vertically to obtain the best fit for the TOP stars), although the models would then fail to predict the observed abundances on the RGB plateau as is the case for NGC\,6397. Whether this discrepancy is due to a lack in our understanding of the first dredge-up at the metal-poor end or a problem with the derived abundances is difficult to say. More research on both fronts, theory and observations, will have to be conducted to understand the underlying physics.\\

The lower estimate to the initial Li abundance based on the T6.00 model does not fully agree with the latest predictions from BBNS by \citet{Cyburt2015} of $2.66\pm0.06$, but is significantly closer than the uncorrected Li abundances. The value agrees well with the Li abundances derived for other metal-poor globular clusters such as NGC\,6397 \citep[$A(\text{Li})=2.46\pm0.09$ at \FeH\ $=-2.10$,][]{Lind2009}, NGC\,6752 \citep[$A(\text{Li})=2.53\pm0.10$ at \FeH\ $=-1.60$,][]{Gruyters2014}, and M4 ($A(\text{Li})=2.57\pm0.10$ at \FeH\ $=-1.10$, Gruyters et al., submitted) \citep{GruytersM4}. We note that these Li values are derived from predictions by the stellar structure model including AD and AddMix, which represents the observations best, and thus from models with different efficiencies of AddMix. The observations seem to suggest that the efficiency of AddMix increases with increasing metallicity. To date, no physical theory has been presented that can explain this relation. Nevertheless, this finding presents theorists with valuable constraints on the physical origin of AddMix.

\subsection{Li as tracer for multiple populations?}
\begin{figure}
\begin{center}
\includegraphics[width=1\columnwidth]{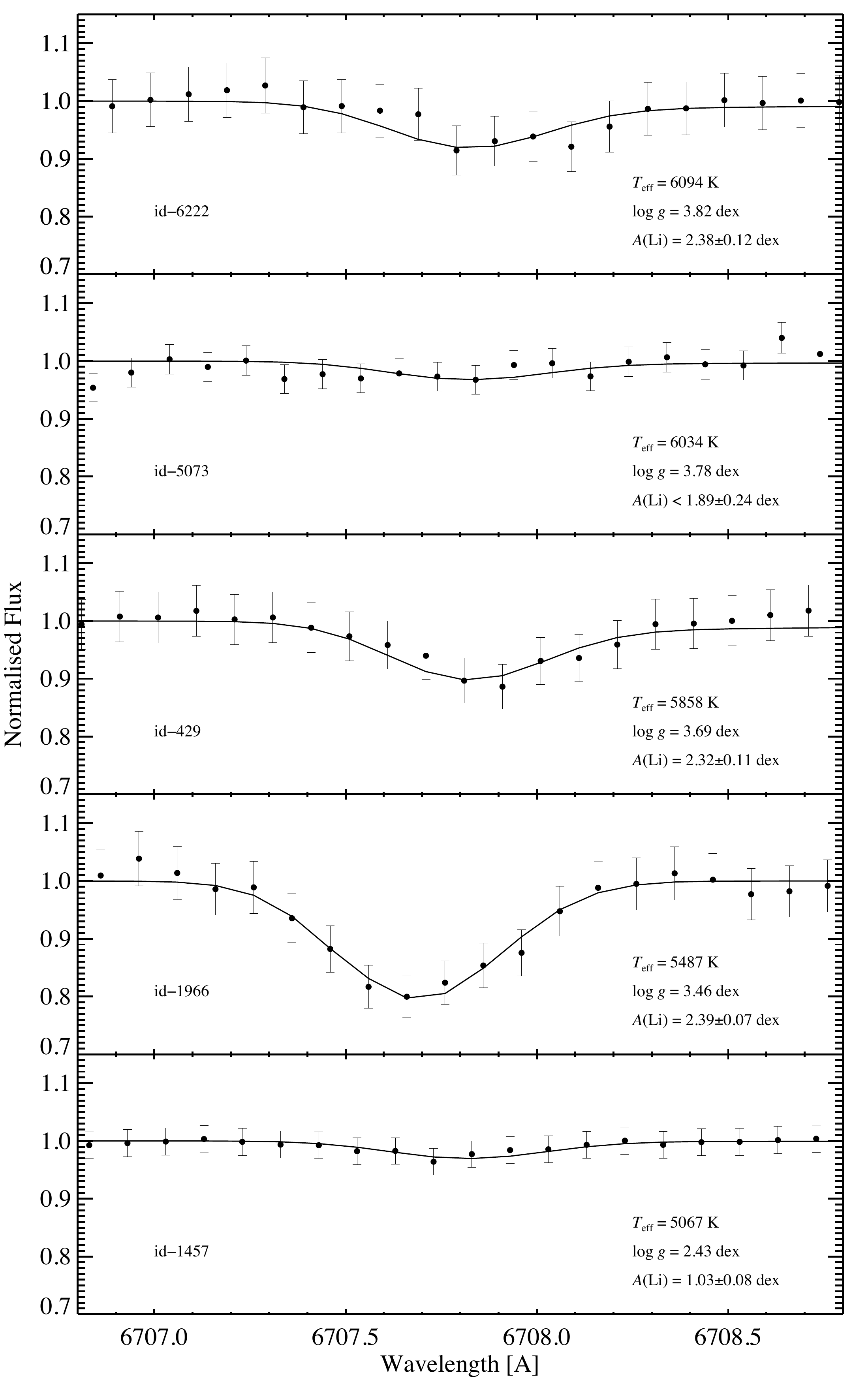}
\caption{Example fits of the Li\,{\scriptsize I} 6707\AA\ line. Each panel shows the stellar id and stellar parameters. From top to bottom we have a Li-normal TOP star, a Li-poor TOP star, a Li-normal SGB star just before Li dilution as a result of the first dredge-up sets in, the Li-rich bRGB star, and a Li-normal RGB star.
}\label{Fig:Li-spectra}
\end{center}
\end{figure}

Figure\,\ref{Fig:Li diffusion} reveals a large spread in derived lithium abundances in the individual evolutionary phases (TOP and RGB). Whether or not the spread is intrinsic is hard to prove given the low data quality on the TOP on the one hand and the weakness of the Li feature on the RGB on the other. We seem to observe four stars pre-first dredge-up ($M>3.25$) and two stars post-first dredge-up ($3>M>1$) that are clearly Li deficient compared to the other stars within these groups. Typical line profiles of the Li 6707\AA\ doublet in stellar spectra of different stellar evolution phases are shown in Fig.\,\ref{Fig:Li-spectra}. The top two panels show the Li line profiles for two stars with very different Li abundances but similar stellar parameters. The spread in Li seems to suggest that the cluster hosts more than one stellar population, as is common in globular clusters. \citet{Lind2009} showed the existence of a Li-Na anticorrelation in the metal-poor globular cluster NGC\,6397. The anticorrelation is a direct result of the past intra-cluster pollution occurring during the early stages of the globular cluster formation. Unfortunately, we do not have access to other abundances to confirm such a Li-Na anticorrelation or any other anticorrelation that may reveal information about a possible pollution epoch in this cluster. Evidence for multiple populations in M30 has been presented by \citet{Monelli2013} and \citet{Piotto2015} using photometry by noting multiple main sequences and a split up of the RGB in the CMD of M30. Earlier, \citet{Carretta2009a} showed the presence of an Na-O anticorrelation in M30. \\

These findings together with our observed spread in Li on both the dwarf and giant plateau strengthens our view that part of the stars in M30 were formed from Li-depleted gas as a result of pollution that occurred early on in the formation of M30. This is fully consistent with the fact that the ejecta of the GC polluters that are responsible for the O-Na anticorrelation have been processed at very high temperature and are thus Li
free. Therefore, second-generation stars that formed out of Li-depleted material mixed with pristine gas were born with an initial Li abundance lower than the BBN value. We thus conclude that the most Li-deficient stars belong to a second generation, formed from intra-cluster gas polluted by a first generation of more massive, faster evolving stars. Our results can be used to evaluate the amount of dilution between pristine material and polluter ejecta (\citealt{Lind2011a} and \citealt{Chantereau2015} and references therein). We considered only TOP stars (\Teff\ $>5900$\,K) with Li abundance higher than $A(\text{Li}) =2.0$ to identify the most likely first-generation stars. Hence we disregarded seven stars of the pre-first dredge-up. The dispersion in this group is rather low, $\sigma(\text{Li})=0.12$, and does not follow any obvious trend with effective temperature or visual magnitude. The average measurement error in abundance stemming from photon noise is 0.10\,dex for the hottest stars. To this we need to add the propagated uncertainty in stellar parameters. When we
include a typical error in effective temperature (50-100\,K), which is the most influential stellar parameter, we can explain a spread in Li abundances of 0.10-0.12\,dex, and thus we conclude that the observed Li abundances on the TOP and RGB plateau are compatible with zero scatter among the first-generation stars. \\
The average Li abundances on the TOP plateau is $2.21\pm0.12$. If we repeat this for the RGB plateau by considering all stars with $5000<$ \Teff\ $<5250$\,K and $A(\text{Li})>1.0,$ we find $1.10\pm0.06$.

\subsection{A Li-rich subgiant}
Among our Li abundances we found one SGB star (id-1966) that is heavily enriched in lithium. With a Li abundance of $2.39\pm0.07$\,dex it resides well above the average Li abundance for giants and dwarfs in M30. Although we only have one observation, we did not detect any anomalies in the stellar radial velocity, and its radial velocity $v_\text{rad} = -186.3 \pm5.0$\,\kms is consistent with that of the cluster, $v_\text{rad} = -183.6 \pm0.6$\,\kms.  
 As we do not detect any abnormalities with the stellar photometry or stellar parameters, we confirm cluster membership of the Li-rich subgiant to M30. The stellar line profile of the Li 6707\AA\ doublet is shown in the fourth panel from the top in Fig.\,\ref{Fig:Li-spectra} and leaves little doubt that the star is indeed lithium rich. It is the strongest profile observed in our sample.\\

\begin{table*}
\caption{Li-rich stars discovered in M30.}
\label{Tab:Li-rich}
\centering
\begin{tabular}{lcccccccccc}\hline\hline
ID & RA  & Dec & $V$ & $V-I$ & $M_V$ & \Teff & $\log g$  & \FeH & $A$(Li) & Ref  \\ 
\# & (J2000) & (J2000) &            &       &      & [K] & [cm\,s$^{-2}$] & [dex] & [dex] & \\
 \hline
1966 & 21 40 09.52 & -23 09 46.6 & 17.739 & 0.806 & 3.01 & 5487 & 3.46 & $\leqslant-2.45\tablefootmark{a}$ & $2.39\pm0.07$ & this work \\
M30-132  & 21 40 09.50 & -23 09 46.4 & 17.60 & 0.72 & 3.04 & 5640 & 3.54 & $-2.43\pm0.12$ & $2.66\pm0.14$ & K16\tablefootmark{b}  \\
M30-7229 & 21 40 18.77 & -23 13 40.4 & 17.05 & 0.75 & 2.49 & 5510 & 3.28 & $-2.32\pm0.11$ & $2.87\pm0.13$ & K16\tablefootmark{b} \\
\hline
\end{tabular}
\tablefoot{The ids for the \citet{Kirby2016} stars refer to the \citet{Sandquist1999} catalogue. Crossmatching our catalogue with the \citet{Sandquist1999} catalogue identified stars M30-132 and 1966 as the same.
\tablefoottext{a}{As a result of the limiting signal-to-noise ratio of the stellar spectrum only an upper estimate for the Fe abundance could be derived.}
\tablefoottext{b}{\citet{Kirby2016}}
}
\end{table*}

Two Li-rich stars were discovered in M30 by \citet{Kirby2016}. The Li-rich star in our sample corresponds to the star with id M30-132 in the sample of \citet{Kirby2016}. The Li enhancement we derive for the star is 0.27\,dex lower than derived by \citet{Kirby2016} but can fully be explained by the difference in stellar parameters (see Table\,\ref{Tab:Li-rich}) and the low signal-to-noise ratio of our spectrum, which does not allow us to derive accurate abundances besides Li. The detection of Li-rich stars in M30 adds to the ever-growing list of globular clusters that host Li-rich stars. \citet{Monaco2012} reported the detection of a Li-rich dwarf in M4 with a Li abundance compatible with the predicted primordial lithium abundance based on standard BBN \citep{Coc2013}. In addition
to M4, there are at least three other globular clusters that seem to harbour stars with Li abundances similar to the primordial Li abundance, NGC\,6752 \citep{Shen2010} and 47\,Tuc \citep{Dorazi2010}, or well above the primordial Li abundance, NGC\,6397 \citep{Koch2011}.\\
Li-rich giants have mostly been detected in the field \citep{Casey2016} \citep[see e.g.][and Casey et al. 2016, submitted  ]{Charbonnel2000,Ruchti2011,Kumar2009,Kumar2011,Lebzelter2012}, and a new study reported that about 0.56\% of the red giants are Li-rich \citep{Zhang2015}. To date, ten Li-rich giants are known in seven globular clusters. These clusters are M3 \citep{Kraft1999}, M5 \citep{Carney1998}, M30 \citep{Kirby2016}, M68 \citep{Ruchti2011,Kirby2016}, NGC\,362 \citep{Smith1999,Dorazi2015}, NGC\,5053 \citep{Kirby2016}, and NGC\,5897 \citep{Kirby2016}. While until recently, only one giant had been found to reside close to the RGB bump \citep{Dorazi2015}, \citet{Kirby2016} now added four more Li-rich stars that have not yet evolved beyond the second dredge-up. Our Li-rich star in M30 is the least evolved star ($V = 17.7$) to date and forms an important data point in bridging the Li evolution from Li-rich dwarfs to Li-rich giants.  Unfortunately, the spectral coverage does not allow us to precisely determine any other abundances, but there is no indication for Ba enhancement, and we thus conclude that the star is not enhanced in the s-process. The lack of s-process enrichment then demonstrates that the Li enrichment comes from mass transfer of an AGB star \emph{before} the thermal pulses set in and can pollute the stellar winds with s-process material. Furthermore, as the star has not reached the RGB and the Li abundance
and after correcting for atomic diffusion ($A({\text{Li}})_{\text{init}} = 2.76$) is comparable to the Li-rich dwarfs detected in other globular clusters, it is reasonable to assume that this star is more closely related to the Li-rich dwarfs than to the Li-rich giants. Clearly, it cannot have produced the large amount of lithium itself. In fact, comparing the Li abundances of all less evolved giants ($M_V > -1)$, we might argue that they are the evolutionary successors of the Li-rich dwarfs because none of them could have produced the large amount of lithium themselves. For a more elaborate discussion on why the mass-transfer scenario is a viable explanation for these stars, see \citet{Kirby2016}. Other scenarios for the Li production in giant stars can be found in \citet{Casey2016}.

\section{Summary}
We have performed a chemical abundance analysis based on GIRAFFE HR15N data of 144 stars in the globular cluster M30 at a metallicity of $-2.3$. We presented NLTE abundances for Ca, Fe, and Li for stars in different evolutionary phases from the TOP to the tip of the RGB.  This is the first time that chemical abundances for the faint TOP in this cluster were derived.  We observed an abundance trend in iron with respect to effective temperature. This trend and the results for Li and Ca can be explained as a result of atomic diffusion reduced by a competing transport
or mixing process referred to as additional mixing.
This is the fourth cluster after NGC\,6397, NGC\,6752, and M4 that shows atomic diffusion signatures. Given the low signal-to-noise ratio on the faint end of the sample, we were unable to distinguish between atomic diffusion models with different additional mixing efficiencies, although the data seem to indicate a similar or lower efficiency than for the globular cluster NGC\,6397 at a metallicity of $-2.1$. An attempt will be made to collect more data in the upcoming observing period to establish the additional mixing efficiency needed to explain the diffusion trends observed in this cluster.\\
We detected one Li-rich subgiant in our sample. With this detection we show that Li-rich stars can be found in all evolutionary phases in globular clusters and thus that Li enhancement probably is not created during the RGB bump phase. Instead, the enhancement in Li could be the result of Li-rich mass transfer during the life cycle of the star. Alternatively, the star could have been formed from locally polluted Li-rich gas. 
The Li abundances for the other stars show the typical Li evolution detected in globular clusters in which the TOP and SGB stars form a thin Li plateau on the same level as the Spite plateau for field stars. Further on in the evolution, the Li abundance drops dramatically as a result of mixing processes connected with low-mass stellar evolution.\\
The spread in Li abundance on the plateau seems to indicate the presence of different stellar populations, but to conclude on
this, other light element abundances need to be derived for the stars. Unfortunately, given the spectral information at hand, this was not possible. The Li plateau value we derived for the dwarfs, after correcting it for atomic diffusion and a conservative efficiency of additional mixing (T6.0 model), is $2.48\pm0.10$ and falls slightly below the agreement window for the primordial Li abundance as derived from the baryonic density deduced from the fluctuations of the cosmic microwave background and standard Big Bang nucleo\-synthesis. The value is the lowest in the series of papers about atomic diffusion in globular clusters, but is still consistent with the diffusion-corrected Li abundances in the other clusters. The lower value could be a result of the different temperature scale used in this work. Steps have been taken to obtain new Str\"omgren photometry for the cluster, and a new dataset should be obtained in September 2016. After it
is properly calibrated, a new temperature scale can be derived that is closely related to the temperature scales used to derive the abundances in the other clusters.

\begin{acknowledgements}
PG and AK thank the European Science Foundation (ESF) for support in the framework of EuroGENESIS. AK acknowledges support by the Swedish National Space Board. O.R. acknowledges HPC at LR and Calcul Qu\'ebec for providing the computational resources required for the stellar evolutionary computations. O.R. also acknowledges the financial support of Programme National de Physique Stellaire (PNPS) of CNRS/INSU. CC acknowledges support from the Swiss National Science Foundation (SNF) for the project 200020-159543 "Multipe stellar populations in massive star clusters. Formation, evolution, dynamics, impact on galactic evolution".
\end{acknowledgements}

\bibliographystyle{aa}
\bibliography{allreferences_M30}
\begin{appendix} 
\section{Stellar parameters and abundances for the group averages}
\begin{table*}
\caption{Derived elemental abundances for the coadded group-averaged spectra.}
\label{Tab:coadd-results}
\centering
\begin{tabular}{lccc cc }
\hline\hline
 Group &  \Teff\ (K) & $\log g$ (cgs) & $\abund{Li}$     & $\abund{Ca}$     & $\abund{Fe}$  \\
 \hline 
Tip-RGB1 & $4288$ & $ 0.80$ &  $-0.51\pm 0.03$ & $ 4.33\pm 0.01$ & $ 5.18\pm 0.03$  \\
Tip-RGB2 &$  4525$ & $ 1.16$ &  $ 0.06\pm 0.05$ &  $ 4.33\pm 0.02$ &  $ 5.16\pm 0.05$  \\
RGB1 & $  4711$ & $ 1.76$ &  $ 0.46\pm 0.11$ &  $ 4.35\pm 0.03$ & $ 5.17\pm 0.04$  \\
RGB2 & $  5031$ & $ 2.35$ &  $ 1.00\pm 0.02$ &  $ 4.36\pm 0.02$ & $ 5.17\pm 0.02$  \\
bRGB1 & $  5175$ & $ 2.76$ &  $ 1.04\pm 0.04$ &  $ 4.33\pm 0.03$ & $ 5.16\pm 0.04$  \\
bRGB2 & $  5337$ & $ 3.19$ & $ 1.17\pm 0.03$ & $ 4.30\pm 0.02$ &  $ 5.16\pm 0.03$  \\
bRGB3 & $  5418$ & $ 3.38$ &  $ 1.30\pm 0.06$ &  $ 4.29\pm 0.05$ & $ 5.11\pm 0.05$  \\
SGB1   & $  5537$ & $ 3.46$ &  $ 1.69\pm 0.05$ &   $ 4.29\pm 0.06$ & $ 5.07\pm 0.07$  \\
SGB2   & $ 5676$ & $ 3.62$ &  $ 1.98\pm 0.03$ &   $ 4.26\pm 0.06$ &  $ 5.03\pm 0.06$  \\
SGB3   & $5822$ &  $ 3.72$ &  $ 2.06\pm 0.03$ &  $ 4.23\pm 0.07$ & $ 5.04\pm 0.07$  \\
TOP1   & $5970$ & $ 3.78$ &  $ 2.00\pm 0.04$ & $ 4.27\pm 0.07$ & $ 5.03\pm 0.07$  \\
TOP2   & $6069$ & $ 3.87$ &  $ 2.00\pm 0.02$ & $ 4.27\pm 0.04$ & $ 4.96\pm 0.04$  \\
TOP3   & $6316$ &  $ 3.99$ & $ 2.15\pm 0.03$ & $ 4.37\pm 0.05$ & $ 4.87\pm 0.06$  \\
\hline
$\Delta \log\epsilon(X)$\tablefootmark{a} & $1247$ & $ 1.83$ & $1.05\pm0.04$ & $-0.05\pm0.06$  & $-0.22\pm0.06$  \\
\hline
\end{tabular}
\tablefoot{Abundance uncertainties are based on the statistical error as calculated by SME.
\\
\tablefoottext{a}{The abundance difference between TOP and RGB. The TOP and RGB abundances are the averages of results for the respective coadded group-averaged spectra.
The uncertainty on the trends is based on the standard deviation of the two averages.
}
\tablefoottext{b}{For the Li trend we take the RGB2 abundance as the cooler RGB group-averaged abundances are affected by the deep-mixing episode occurring on the RGB.}
\\
\tablefoottext{Li}{Based on Li\,{\scriptsize I} $\lambda$6707.8, assuming NLTE.}
\tablefoottext{Ca}{Based on three Ca\,{\scriptsize I} $\lambda\lambda$6493.9, 6499.7 and 6717.7, all assuming NLTE.}
\tablefoottext{Fe}{Based on five Fe\,{\scriptsize I} and one Fe\,{\scriptsize II} lines $\lambda\lambda$6495.0, 6518.4, 6592.9,  6593.9, and 6678, and $\lambda$6516.1, assuming NLTE.}
}
\end{table*}

\end{appendix}

\end{document}